\documentclass[iop]{emulateapj}
\usepackage{epsfig}
\usepackage{apjfonts}
\usepackage{amsmath,color,bm}
\usepackage{booktabs}
\usepackage[breaklinks,colorlinks,urlcolor=blue,citecolor=blue,linkcolor=blue]{hyperref}
\usepackage{natbib}
\usepackage{nameref}
\usepackage{graphicx}
\usepackage{graphics}
\usepackage[space]{grffile}
\usepackage{latexsym}
\usepackage{amsfonts,amsmath,amssymb}
\usepackage{url}
\usepackage[utf8]{inputenc}
\usepackage{fancyref}
\usepackage{hyperref}
\usepackage{xcolor}
\usepackage[normalem]{ulem}

\newcommand{\ignore}[1]{}

\newcommand{\Mej}{M_\mathrm{ej}}

\newcommand{\Lbolmax}{L_\mathrm{bol}^\mathrm{max}}
\newcommand{\dLbol}{\Delta \log L_\mathrm{bol}}

\newcommand{\Msun}{\mathrm{M}_\odot}
\newcommand{\vej}{v_\mathrm{ej}}
\newcommand{\Ye}{Y_\mathrm{e}}

\begin{document}
	
\title{Can Kilonova Light Curves be Standardized?}
	
\author{Rahul Kashyap$^1$, Gayathri Raman$^1$, Parameswaran Ajith$^{1,2}$}
\address{$^1$ International Centre for Theoretical Sciences, Tata Institute of Fundamental Research, Bengaluru 560089, India}
\address{$^2$~Canadian Institute for Advanced Research, CIFAR Azrieli Global Scholar, MaRS Centre, West Tower, 661 University Ave, Toronto, ON M5G 1M1, Canada}
	
\begin {abstract}
Binary neutron star mergers have been recently confirmed to be the progenitors of the optical transients kilonovae (KNe). KNe are powered by the radioactive decay of neutron-rich elements (r-process elements) which are believed to be the product of disruption of neutron stars during their merger. KNe exhibit interesting parallels with type Ia supernovae (SNe), whose light curves show specific correlations which allow them to be used as standardizable candles. In this paper, we investigate whether KNe light curves could exhibit similar correlations.  While a satisfactory answer to this question can only be provided by future KNe observations, employing theoretical models we explore whether there is any ground for harboring such expectations. Using semi-analytic models of KNe light curves in conjunction with results from numerical relativity simulations of binary neutron star mergers, we obtain the maximum bolometric luminosity ($L_{\mathrm{Bol}}^{\mathrm{max}}$) and decline from peak luminosity ($\Delta \log L_{\mathrm{Bol}}$) for a simulated population of mergers. We find that theoretical light curves of KNe show remarkable correlations despite the complex physics governing their behavior. This presents a possibility of future observations to uncover such correlations in the observed light curves, eventually allowing observers to standardize these light curves and to use them for local distance measurements. 
\end {abstract}
	
\keywords{kilonovae, supernovae: general --- neutron stars, compact binary coalescence, standardizable candle}

\section{Introduction}
	
The gravitational wave (GW) event GW170817 marked the birth of a new era in multi-messenger astrophysics~\citep{TheLIGOScientific:2017qsa}. The event was associated to a binary neutron star (BNS) merger located nearly 40 Mpc away in the galaxy NGC\,4993. The GW trigger was followed by a nearly coincident short gamma ray burst (sGRB) 1.7\,s after the merger time~\citep{Monitor:2017mdv}. The gamma-ray counterpart was subsequently followed up by several ground and space based telescopes in the ultraviolet, optical and near-infrared (hereafter, UVOIR) bands~\citep{Abbott2017A}. The UVOIR emission is largely identified to be from a quasi-thermal transient called a kilonova (KN), powered by radioactive decay of several r-process nuclei \citep{LiPaczynski1998,Metzger2010,Barnes2013A}, resulting in typical luminosities of $\sim$10$^{40}$ - 10$^{42}$ erg s$^{-1}$.   

A detailed description of BNS mergers has been obtained using large scale numerical relativity (NR) simulations by various groups (see, e.g., \citealt{shibata_kenta2019,Tanaka2016,Rosswog2015IJMPD,FaberRasio2012} and references therein). These groups have found that the properties (e.g., masses and velocities) of r-process radioactive material ejected from the merger tends to exhibit a non-trivial dependency on the neutron star (NS) masses and tidal deformabilities. Various groups have provided formulae that gives excellent fits to the ejecta {properties} as a function of the NS masses and equation of state (EOS) \citep{Radice2018A,Coughlin2019}. Radiative transfer calculations performed using data from such numerical simulations obtain light curves that qualitatively agree with the observations of GW170817 \citep{Tanaka2017A, Kasen2017, Tanvir_2017, Miller2019}.
Additionally, semi-analytical models based on Arnett-Chatzopoulos framework \citep{Arnett1982A, Chatzopoulos2012} (originally developed for Type Ia supernovae) have been able to explain the light curve fairly well. In such models, the heating rate is calculated using radioactive decay of mostly r-process elements,  giving the initial intensity decline rate to be $t^{-1}$ followed by $t^{-3}$~\citep{Arnett1982A, LiPaczynski1998, Metzger2010, Chatzopoulos2012,Korobkin2012, Metzger2017, Barnes2016A, Villar2017A, Arcavi2017A}. 

It is interesting to note how much of this picture of the electromagnetic transient described above is similar to that of type Ia supernovae (SNe Ia). SNe Ia are bright optical transients (typical luminosities $\sim$10$^{43}$ - 10$^{44}$erg s$^{-1}$) formed from thermonuclear explosion of white dwarfs \citep{Hoyle-Fowler1960A,Wheeler1990A} whose light curves are powered by radioactive decay of Ni$^{56}$. The double degenerate model proposes binary white dwarf mergers as a progenitors of SNe Ia \citep{Iben1984A, Webbink_1984A, Nelemans2001A}. This model is being supported by several recent observational and theoretical studies, particularly in terms of it being able to explain the Galactic birth rates and delay time distributions \citep{Ruiter2009A,Maozetal_2013,Kashyap2015A}.

It is well known that SNe Ia light curves show an empirical relationship between the maximum intrinsic luminosity and the decline rate, known as the ``Phillips relation'' or the ``width-luminosity relation'' \citep{Phillips1993A}. By estimating the maximum intrinsic luminosity from the observed decline rate using the Phillips relation they can be used as a standard candle \citep{Riess1998A}. There are several parallels between SNe Ia and KNe: Both are believed to be triggered by the merger of compact objects in narrow mass ranges and are powered by radioactive decay of heavy isotopes. Moreover empirical models based on Arnett-Chatzopoulos framework seem to agree with the observations. Hence, it is quite natural to ask this question: Could KNe light curves be standardized like SNe Ia light curves? 

A definitive answer to this question can only be provided by a large number of KNe observations, as happened in the case of SNe Ia. As we wait for such observations~\citep{Yang2017}, we explore whether there is any ground for harboring such expectations. This is done by investigating whether any correlations exist in the synthetic KNe light curves provided by semi-analytical models in conjunction with results from NR simulations of BNSs. Making use of NR fitting formulae for the ejecta properties, we generate synthetic light curves from several putative KNe produced by the merger of several simulated BNS systems with different component masses. We then investigate the correlation between the peak luminosity ($L_{\mathrm{Bol}}^{\mathrm{max}}$) and the decline in luminosity ($\Delta \log L_{\mathrm{Bol}}$)  after 5 days following the peak. We find that ``Phillips-like'' relations exist in these synthetic light curves. 

Indeed, the current semi-analytical models are unlikely to capture the complex physics and the rich phenomenology of KNe in entirety. Hence, the specific relationship that we find using the current semi-analytic models are unlikely to hold up against actual observations. However, they hint a possibility of the existence of such relationships in real light curves. This paper is organized as follows: Sec.~\ref{KN_analytic} provides a summary of synthetic light curve models that we are employing along with the NR fitting formulas for ejecta properties. In Sec.~\ref{results} we  discuss  our main results while Sec. ~\ref{sec:outlook} presents a summary and outlook.

\section{Semi-analytical modeling of Kilonova light curves}
\label{KN_analytic}

\begin{figure}
	\centering
	\includegraphics[width=3.45in]{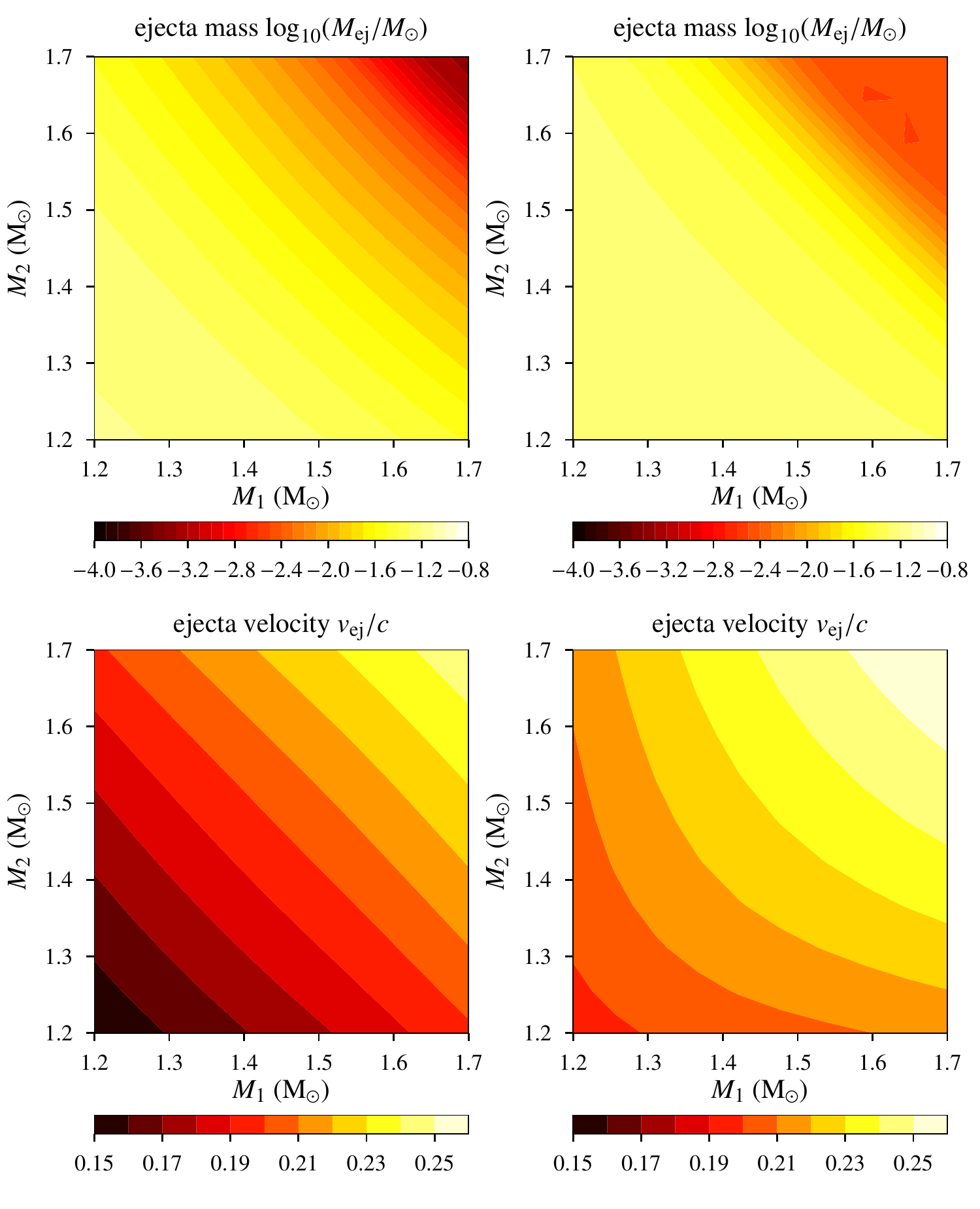}
	\caption{Total ejecta mass $\log_{10}(\Mej/M_{\odot})$ (top panels) and ejecta velocity $\vej/c$ (bottom panels) as a function of the NS masses $M_1$ and $M_2$ in the binary, computed using the fits given in \cite{Radice2018A} (left panels) and \cite{Coughlin2019} ({right panels}). We assume here that 30\% of the disk mass contributes to the unbound r-process ejecta that powers the light curve.} 
	\label{ejecta_fit}
\end{figure}

\begin{figure}
	\centering
	\includegraphics[width=3.3in]{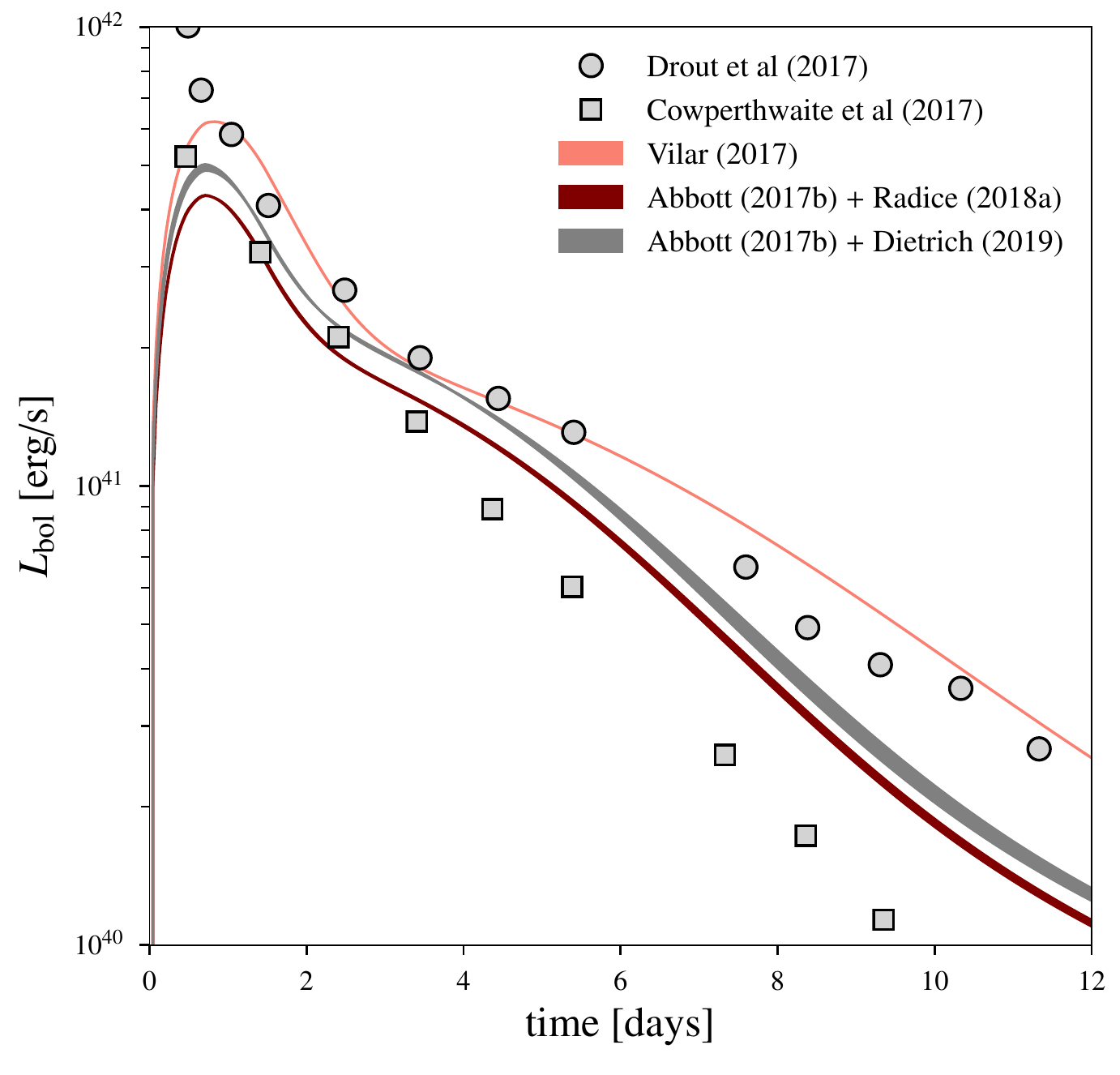}
	\caption{The solid traces show the evolution of bolometric luminosity of the KNe associated with GW170817 as predicted by the semi-analytical KNe models. The Different markers show the observed luminosity evolution from the same event calculated by \cite{Drout2017} and \cite{Cowperthwaite2017}.}
	\label{bolo-lc}
\end{figure}

In the absence of a large enough number of KNe observations, here we explore the possibility of the existence of a {Phillips}-like relation in the synthetic light curves predicted by semi-analytical KNe models. 
In this paper we adopt the Arnett-Chatzopoulos model to obtain the light curves for a population of binary NS mergers using the NR fitting formulae for ejecta mass and velocity provided by \cite{Radice2018A} and \cite{Coughlin2019}. 

The light curve modeling justifiably assumes that the physical processes responsible for heating that produces UVOIR are well separated in time from the processes responsible for $\gamma$-rays, X-rays and radio \citep{Kenta2016}. In addition it is assumed that there is a homologously expanding isotropic ejecta of neutron-rich radioactive isotopes. This expanding ejecta will follow the same evolution in an ambient medium as seen for SNe, for example. The model constructs the light curve by taking into account the work done in expansion, the heating done by radioactive decay along with the knowledge of velocity and opacity of the expanding ejecta \citep{Arnett1982A, Chatzopoulos2012}. The properties of the ejecta, in turn, depend on the mass of the NSs in the binary, for a given EOS as shown by NR simulations~\citep{Dietrich2017C,Radice2018A,Coughlin2019}.

Ejecta masses and velocities are functions of gravitational and baryonic masses, mass-ratio and weighted average $\tilde{\Lambda}$ of individual tidal deformabilities \citep{Flanagan:2007ix}. From the NS masses in the binary, assuming the {DD2 EOS  \citep{DD22010-1,DD22010-2}}, we obtain the masses and velocities of the dynamical ejecta and disk mass using NR fits provided by \cite{Radice2018A} (Eqn 18-25) and \cite{Coughlin2019} (Eqn D1-D5). We assume that 30\% of the disk mass becomes unbound and contributes as the disk  ejecta. The total ejecta mass, $\Mej$ is then taken to be the sum of the dynamical and disk ejecta. Figure~\ref{ejecta_fit} shows the ejecta mass and velocity as a function of the NS masses in the binary. We find that the numerical fits of the eject mass (velocity) provided by both groups differ, on average, by $\sim$ 29\% (12\%), which is a reflection of the error in these estimates. The NS radius decreases with the mass; hence, the lower-mass companion is more prone to tidal deformation, producing larger ejecta mass. Also, since more massive NSs are more compact, they would also produce larger ejecta velocities. We observe these trends in figure \ref{ejecta_fit}.

\begin{figure*}
	\centering 
	\includegraphics[width=6.3in]{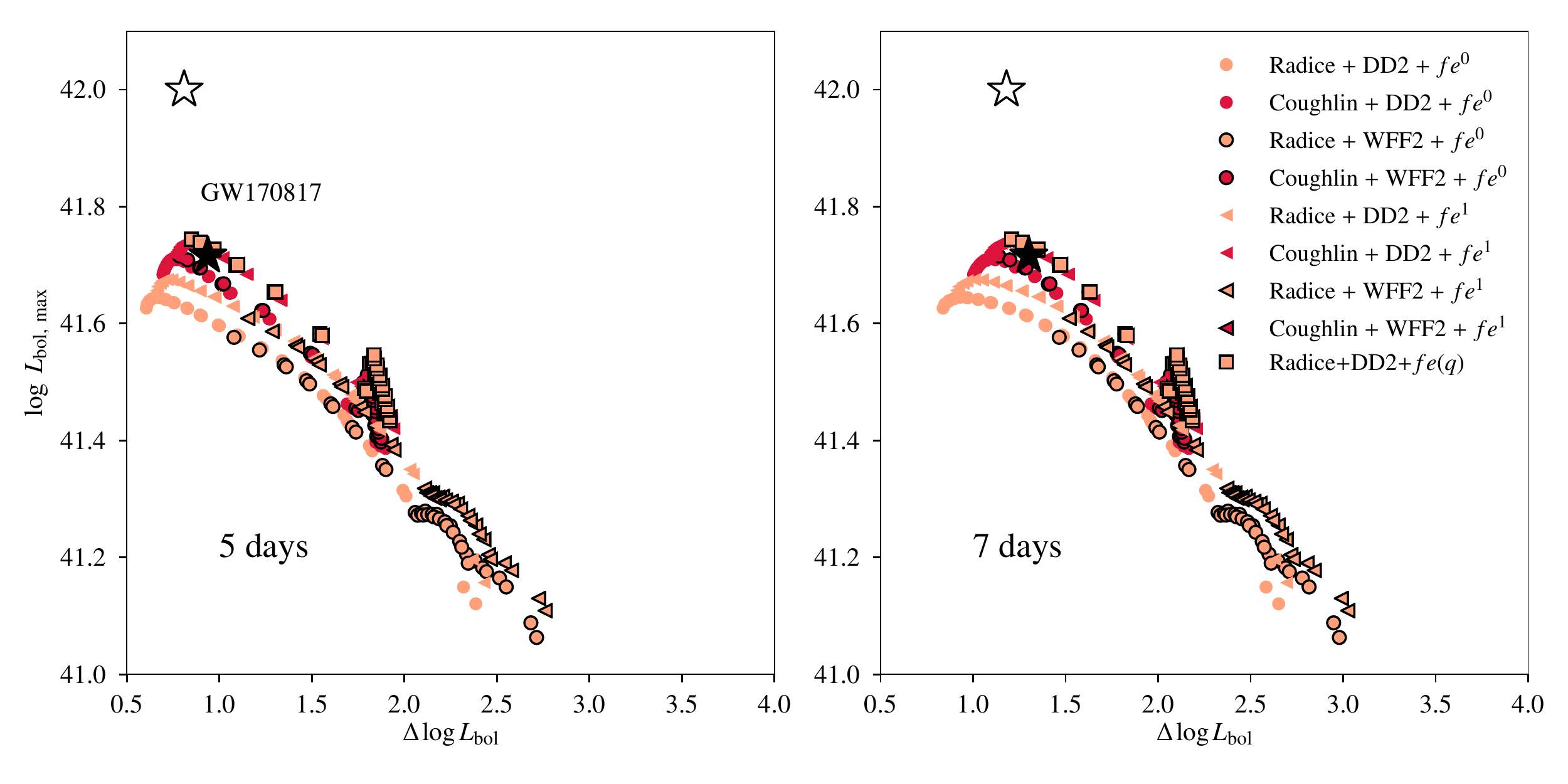}
	\caption{The relation between maximum luminosity $\Lbolmax$ and decline in luminosity in few days $\dLbol$ from simulated light curves. We consider masses in the range $1.2-1.7~M_\odot$, two different NR fitting formulae for ejecta mass and velocity (\cite{Radice2018A} and \cite{Coughlin2019}; shown in different colors), two different EoSs (DD2 and WFF2; shown by markers with and without black edges), and two different choices of ${f_{\rm e}}$ (${f_{\rm e}}^0 \equiv [0.2, 0.6, 0.2]$, ${f_{\rm e}}^1 \equiv [0.26, 0.6, 0.14]$ and the mass-ratio dependent ${f_{\rm e}}(q)$ as discussed in the text; shown by different markers). The filled and unfilled stars correspond to the GW170809 KNe observations by \cite{Cowperthwaite2017} and \cite{Drout2017}. The left panel corresponds to a delay time of 5 days and the right panel to 7 days.} 
	\label{correlations_lc}
\end{figure*}
\bigskip 

In our model, the total ejecta mass is then decomposed into ``blue'', ``purple'' and ``red'' components, which differ in their electron fraction $\Ye$ and hence the opacity $\kappa$. Following \cite{Villar2017A}, we assume $\kappa_m = 0.5, 3, 10  ~ \mathrm{cm}^2 ~ \mathrm{g}^{-1}$ for the blue, purple and red components, respectively. Finally, we add the light curve due to the three components to obtain bolometric luminosity. We use the symbol ${f_{\rm e}}$ to denote array of the fraction of ejecta mass distributed in to the blue, purple and red components (${f_{\rm e}}_m$ being $m$th element of this array): for example, ${f_{\rm e}}=[0.2,0.6,0.2]$ means that total ejecta mass is decomposed into 20\% blue ($\kappa= 0.5~\mathrm{cm}^2 ~ \mathrm{g}^{-1},~\Ye > 0.4$), 60\% purple ($\kappa = 3~\mathrm{cm}^2 ~ \mathrm{g}^{-1},~ 0.1 < \Ye < 0.4$) and 20\% red ($\kappa = 10~\mathrm{cm}^2 ~ \mathrm{g}^{-1}, \Ye < 0.1$) components. Note that ${f_{\rm e}}$ of dynamical and disk ejecta can be, in general, different~(see, e.g., \cite{Radice2018A} and \cite{Lippuner:2017bfm}). Leaving a more careful treatment of this for a future work, we consider three simple choices of ${f_{\rm e}}$ (see section~\ref{results}). In the absence of accurate predictions from NR simulations, we assume the same velocity $\vej$ for all components of the ejecta.

The analytical modelling of light curve depends on the input heating rate and thermal efficiency of each component of the ejecta, given by \cite{Korobkin2012} and \cite{Barnes2016A}, respectively.
\begin{eqnarray}\label{eq:L_bol}
L_{\mathrm{in},~m} (t) & = & 4\times 10^{18} \ M_{\mathrm{rp},~m}\ [0.5 - \pi^{-1} \arctan \bigg(\frac{t-t_o}{\sigma}\bigg)]^{1.3} \rm \ erg\ s^{-1} \nonumber \\ 
\epsilon (t) & = & 0.36\bigg[\exp({-a \, t})+\frac{\ln(1 + 2 \, b \, t^{d})}{2 \, b \, t^{d}}\bigg] 
\end{eqnarray}
where $M_{\mathrm{rp},\,m}$ is the total mass (in $\Msun$) of the r-process elements synthesized for each component $m$ (blue, purple or red), i.e., $M_{\mathrm{rp},~m} = {f_{\rm e}}_m \, \Mej$, and $t$ is time in days. We use 2-d interpolation for each of the ejecta components to obtain the values for the fit parameters $a, b, d$ for different ejecta masses and velocities using the Table 1 of \cite{Barnes2016A}. 

Assuming homologous expansion as described in \cite{Arnett1982A}, we use the prescription outlined in \cite{Chatzopoulos2012} and \cite{Villar2017A} to compute the luminosity for each component $m$  	
\begin{equation}
L_m(t) = \exp\left(\frac{-t^2}{\tau_m ^2}\right) \int_0 ^t 2 \, L_{\mathrm{in},\,m}(t') \, \epsilon(t') \, \exp\left(\frac{t'^2}{\tau_m ^2}\right) \, \frac{t'}{\tau_m^2} \, dt'
\end{equation}
where, $\tau_m = \sqrt{{2\kappa_m M_{\mathrm{rp},\,m}} / {\beta \vej  c}} $, with $\kappa_m$ being is the gray opacity of the ejecta component, and $\beta = 13.4$, a dimensionless constant associated with the geometric profile of the ejecta~\citep{Villar2017A}. 

Figure~\ref{bolo-lc} shows the synthetic light curves computed for the KN associated with GW170817, along with the observed ones. \cite{Villar2017A}'s best fit model uses $M_{\mathrm{rp},~m} = [0.02, 0.047, 0.011] \mathrm{M}_\odot$ and $\vej\,_{,\,m} = [0.266, 0.152, 0.137] \, c$ for the blue, purple and red components of the ejecta.  We also plot the light curves computed using the estimated component masses from GW170817 ($ M_1 = 1.36 - 1.6 \mathrm{M}_\odot$; $M_2 = 1.16 - 1.36 \mathrm{M}_\odot$, with the constraint $M_1 + M_2 = 2.73 ^{+0.04} _{-0.01} M_{\odot} $~\citealt{TheLIGOScientific:2017qsa}), where the ejecta mass and velocity estimated using NR fitting formulae of \cite{Radice2018A} and \cite{Coughlin2019}. Here, as discussed earlier, we assume that ${f_{\rm e}}=[0.2,0.6,0.2]$. For comparison, we also plot the observed light curves as presented by \cite{Drout2017} and \cite{Cowperthwaite2017}.  The general agreement between the theoretical models and observations is encouraging. 

\section{Results} 
\label{results}

We generate a population of BNS mergers whose masses are uniformly distributed in the mass range $1.2-1.7~\mathrm{M}_\odot$ and compute the synthetic light curves produced by each merger, using the procedure outlined in section~\ref{KN_analytic}. We use these theoretical light curves to find the relation between maximum luminosity $\Lbolmax$ and decrease in luminosity $\dLbol$ in 5 days from the time of the peak luminosity, where $\dLbol \equiv \log (\Lbolmax / L_\mathrm{bol}^\mathrm{5~days})$. The choice of 5 days is arbitrary, but is motivated by the fact that UVOIR observations of KNe can be typically performed over a few days. We vary the parameters in the model and discuss the possible variations in the correlation. In particular, we vary the choice of NR based fitting formula for the ejecta mass and velocity (provided by \citealt{Radice2018A, Coughlin2019}), the nuclear EOS (DD2 by \citealt{DD22010-1} and WFF2 by \citealt{Wiringa1988-WFF2}), and distribution of the electron fraction $\Ye$ of the ejecta (${f_{\rm e}}=[0.2,0.6,0.2]$ and ${f_{\rm e}}=[0.26, 0.6, 0.14]$ used by ~\citealt{Villar2017A}). We also consider a case where ${f_{\rm e}}$ depends on the NS masses\footnote{We use figure 5 from \citet{Dietrich2017C} to get the average electron fraction of the ejecta, $\bar{Y}_e(q)$,  as a function of the mass ratio $q := M_1/M_2$ of the binary. Then we solve two constraint equations, $\sum_m {f_{\rm e}}_m =1$ and $\sum {f_{\rm e}}_m \, {\Ye}_m =\bar{Y}_e(q)$, where $m$ denotes the blue, purple and red components. The system is under-determined as there are three unknowns (${f_{\rm e}}_m$) and only two equations. We choose ${f_{\rm e}}_\mathrm{blue} = 0.1$ for the blue component to obtain the same for the red and purple components. Corresponding results are plotted in figure~\ref{correlations_lc}. A choice ${f_{\rm e}}_\mathrm{blue} = 0.2$ does not make a big difference.}. We also examine the same correlations computed assuming a time delay of 7 days from peak luminosity. These results are plotted in figure~\ref{correlations_lc}, suggesting clear correlations between $\Lbolmax$ and $\dLbol$. 
  
It should be mentioned here that the relation found in this work factors in the full non-linearity in the NR simulations and the non-trivial relationship between NS masses and bolometric light curve. The fact that such a correlation has been observed in the synthetic light curves suggests that a similar correlation should exist in the actual light curves, even though the actual observed correlation may turn out to be different than what are presented here. If this is vindicated by future KNe observations, this will provide an independent distance ladder. For example, in figure \ref{correlations_lc}, the decline in luminosity in 5 days (or any other suitably chosen time) is an independent observable which can be used to find the maximum intrinsic luminosity using the correlation. Thus the luminosity distance can be estimated by comparing the intrinsic and apparent luminosities.

\section{Summary and outlook}
\label{sec:outlook}

Motivated by the similarities of KNe with SNe Ia, we have explored the possibility of KNe providing a set of standardizable candles analogous to SNe Ia. Indeed, such a possibility can be confirmed or refuted only by a large number KNe observations. As we await such observations, we studied simple semi-analytical KNe models (in conjunction with NR fitting formulae for ejecta mass and velocity) and discovered correlations that exist between the peak bolometric luminosity $\Lbolmax$ and the decline in the luminosity $\dLbol$ after a few days (figure~\ref{correlations_lc}). This is performed by computing $\Lbolmax$ and $\dLbol$ from synthetic light curves generated from ejecta produced by a population of BNS mergers. We employ different NR fitting formulae, NS EOS and electron fraction distribution of the ejecta to study the robustness of our results. We note that \cite{Coughlin:2019vtv} has taken a different approach to the standardization of KNe. 

The light curve calculation presented in this work has multiple simplifying assumptions, and are subject to errors in the NR simulations and the KNe models. Hence they are only crude estimates. In particular, anisotropy of the ejecta components and time-variation of the ejecta opacities needs future investigation. However, there is preliminary evidence (coming from the observation of KNe associated with GW170817) that they capture the essential features of KNe light curves. We stress the fact that we are not proposing any particular correlation, which has to be left to future observations. Such a correlation in the observed light curves could have potential usage in distance measurement, which will have applications in fundamental physics, astrophysics and cosmology. Possible applications include constraining the number of spacetime dimensions by comparing distance estimates from GW and KN observation from a BNS merger (along the lines of  \citealt{Pardo:2018ipy,Abbott20199}),  the calibration of distance ladders (along the lines of \citealt{Gupta2019}), and potentially the estimation of cosmological parameters (along the lines of \citealt{Coughlin:2019vtv}).

We admit that there are however key differences between SN and KNe light curves. KNe have peak luminosities ($\sim$10$^{40}$ - 10$^{42}$ erg s$^{-1}$) much lower than SNe Ia peak luminosities ($\sim$10$^{43}$ - 10$^{44}$ erg s$^{-1}$) making KNe observable only in the local universe. The SNe Ia B-band light curve usually peaks $\sim$20 days post explosion \citep{Riess_1999} where the spectral peak shifts from optical to infrared in about 2-3 months. In contrast, the KNe light curve reaches the maximum value within few hours and shifts from optical to infrared within 10 days. Because of these differences, KNe light curves are relatively more difficult to standardize and also the counterpart of the ``Phillips relation'' would be expected to be different for them. Only future observations can tell the full story. 

\paragraph{Acknowledgments:} We thank Kenta Hotokezaka, Bala Iyer, Ramesh Narayan, Sumit Kumar, K.~Haris for their patient feedback to the work. We also thank Bharat Kumar for providing polytropic fits to the EOS DD2, and David Radice and Tim Dietrich for clarifications on their NR fits. RK would like to thank G. Srinivasan for his encouragement to pursue the problem during a meeting in 2017 at ICTS, Bengaluru. We are grateful to the anonymous referee for many useful comments and suggestions on an earlier version of the manuscript. This research was supported by the Max Planck Society through a Max Planck Partner Group at ICTS-TIFR and by the Canadian Institute for Advanced Research through the CIFAR Azrieli Global Scholars program. Computations were performed at the ICTS cluster Alice. 
\bibliography{Binary_NS.bib}
	
\end{document}